\documentclass[10pt,pra,twocolumn,floatfix,showpacs]{revtex4}
\usepackage{amsfonts}
\usepackage{amssymb}
\usepackage{amsmath}
\usepackage[dvips]{graphicx}

\begin{document}

\title{Gates for one-way quantum computation based on
Einstein-Podolsky-Rosen entanglement}
\author{Shuhong Hao, Xiaowei Deng, Xiaolong Su}
\email{suxl@sxu.edu.cn}
\author{Xiaojun Jia, Changde Xie, and Kunchi Peng}
\affiliation{State Key Laboratory of Quantum Optics and Quantum Optics Devices,\\
Institute of Opto-Electronics, Shanxi University, Taiyuan, 030006, People's
Republic of China}

\begin{abstract}
Single-mode squeezing and Fourier transformation operations are two
essential logical gates in continuous-variable quantum computation, which
have been experimentally implemented by means of an optical four-mode
cluster state. In this paper, we present a simpler and more efficient
protocol based on the use of Einstein-Podolsky-Rosen two-mode entangled
states to realize the same operations. The theoretical calculations and the
experimental results demonstrate that the presented scheme not only
decreases the requirement to the resource quantum states at the largest
extent but also enhances significantly the squeezing degree and the fidelity
of the resultant modes under an identical resource condition. That is
because in our system the influence of the excess noises deriving from the
imperfect squeezing of the resource states is degraded. The gate operations
applying two-mode entanglement can be utilized as a basic element in a
future quantum computer involving a large-scale cluster state.
\end{abstract}

\maketitle

\section{Introduction}

Over the past few decades a variety of fundamental protocols for
implementing quantum computation (QC) have been explored \cite%
{Nielsen2000,Loock1}. There are two different models in the QC regime, which
are the traditional circuit model, in which unitary evolution and coherent
control of individual qubits are required \cite{Nielsen2000}, and the
cluster model, in which the logical operations are achieved through
measurements and classical feedforward of measured results on a cluster
entangled state \cite{Raussendorf2001}. Due to the role of measurements the
QC based on cluster entanglement is essentially irreversible, and thus it is
named the one-way QC \cite{Raussendorf2001}. The one-way QC was first
experimentally demonstrated with a four-qubit cluster state of single
photons \cite{Walther2005,Prevedel2007,Chen2007}. In the meanwhile, an
universal QC model using continuous-variable (CV) cluster states was
proposed \cite{Menicucci2006}. Applying the approach of quantum optics, CV
cluster states of optical field can be unconditionally prepared \cite%
{Zhang2006,Loock2007,Su2007,Yukawa2008}, and the one-way CVQC can be
deterministically performed \cite{Menicucci2006,Weedbrook2012}. Therefore,
the probabilistic problems existing in most qubit information systems of
single photons \cite{Walther2005,Prevedel2007,Chen2007} can be overcome. It
has been theoretically and experimentally demonstrated that one-mode linear
unitary Bogoliubov (LUBO) transformations corresponding to Hamiltonians that
are quadratic in quadrature amplitude and phase operators of quantized
optical modes (qumodes) can be implemented using a four-mode linear cluster
state \cite{Ukai2010,Ukai2011}. At the same time, the Deutsch-Jozsa
algorithm for CVQC has been proposed \cite{Zwi}. Following the theoretical
proposals, the different logical gates used for CVQC were experimentally
realized. First, a quantum nondemolition sum gate and a quadratic phase gate
for one-way CVQC were demonstrated based on utilizing squeezed states of
light by Furusawa's group in 2008 and 2009, respectively \cite%
{Yoshikawa2008,Miwa2009}. Successively, a controlled-X gate based on a
four-mode optical CV cluster state was presented by Peng's group, in which a
pair of quantum teleportation elements were used for the transformation of
quantum states from input target and control states to output states \cite%
{Wang2010}. Later, the squeezing operation, Fourier transformation and
controlled-phase gate were also achieved by Ukai et al., in which four-mode
optical cluster states served as resource quantum states \cite%
{Ukai2011,Ukai20112}.

Here, we present a measurement-based logical operation scheme with which the
squeezing and Fourier transformations for a single qumode can be implemented
using an Einstein-Podolsky-Rosen (EPR) entangled state as the resource.
These operations can be achieved on a fixed experimental system only by
choosing appropriate measurement angles in homodyne detections. Since EPR
entanglement of optical modes is deterministic and homodyne detections can
be well controlled, the presented CVQC gates are operated in a completely
unconditional and controllable version. By changing the quadrature
measurement angles of homodyne detections the squeezing operations at three
different squeezing levels ($-4$, $-8$, $-12$ dB) and Fourier transformation
are experimentally performed. The experimental results and the corresponding
theoretical expectations are in good agreement. As is well known, the EPR
entanglement is equivalent to a two-mode cluster state \cite{Zhang2006}, and
thus, QC using an EPR state can be implemented on two submodes of a large
cluster state as a step of a full one-way CVQC. We also prove that the
squeezing degree and the fidelity of the output mode obtained by using an
EPR state are better than that obtained using a four-mode cluster state if
the squeezing of the initial resource state is identical. That is because
the excess noises deriving from imperfect squeezing of the resource state in
the EPR system are less than that in the four-mode cluster state system.
Theorefore, the presented CVQC schemes not only decrease the requirement to
quantum resource and simplify the experimental system significantly but also
enhance the quality of the output states. Finally, we point out that the
set-up can be applied to perform a cascaded operation of a squeezing gate
and a Fourier gate by changing the relative phase between the input mode and
a submode of the EPR state, which shows further the diversity of the
protocol.

\section{Protocol and principle of quantum logical operations}

The single-mode squeezing gate in CVQC depending on quantized optical modes%
\textbf{\ }is expressed by $\hat{S}(r)=e^{ir(\hat{x}\hat{p}+\hat{p}\hat{x})}$%
, where $r$ is the squeezing parameter, $\hat{x}=(\hat{a}+\hat{a}^{\dagger
})/2$ and $\hat{p}=(\hat{a}-\hat{a}^{\dagger })/2i$ are the amplitude and
phase quadratures of an optical mode $\hat{a}$, respectively. The
input-output relation of the squeezing gate is written as $\mathbf{\hat{\xi}}%
{\acute{}}%
_{j}=\mathbf{S\hat{\xi}}_{j}$, where $\mathbf{\hat{\xi}}_{j}=$ $(\hat{x}_{j},%
\hat{p}_{j})^{T}$ and 
\begin{equation}
\mathbf{S=}%
\begin{pmatrix}
e^{r} & 0 \\ 
0 & e^{-r}%
\end{pmatrix}%
\end{equation}%
represents the squeezing operation of the phase quadrature.

\begin{figure}[tbp]
\setlength{\belowcaptionskip}{-3pt} 
\centerline{
\includegraphics[width=80mm]{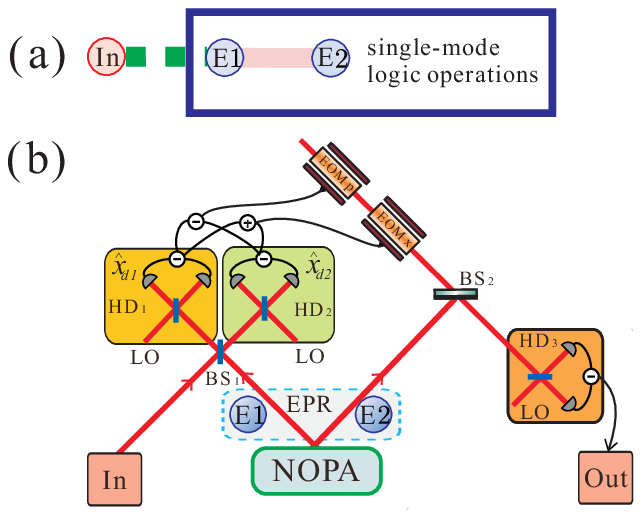}
}
\caption{(Color online) The schematic of a single-mode quantum logic
operation with an EPR entangled state. (a): the graph representation, (b):
experimental set-up. The input state $\protect\alpha $ is coupled to an EPR
entangled state E1-E2 via a $50\%$ beam-splitter BS$_{1}$. Measurement
results from two homodyne detection systems (HD$_{1}$ and HD$_{2}$) are
fedforward to modes E2. The output mode is measured by HD$_{3}$. LO: local
ossilator for the homodyne detection. EOMx and EOMp: amplitude and phase
electro-optical modulators. BS$_{2}$: a mirror with 99\% reflection
coefficient. }
\end{figure}

Fig. 1 shows a schematic of the single-mode squeezing and Fourier
transformation gate based on applying an EPR entangled state, (a) is the
graph representation, (b) is the experimental set-up. An input mode is
coupled to a submode of the EPR entangled state (E1) via a 50\%
beam-splitter BS$_{1}$. The two output modes of BS$_{1}$ are measured by
homodyne detection systems HD$_{1}$ and HD$_{2}$, respectively. The measured
results are fedforward to the other submode of the EPR entangled state (E2)
by classical feedforward circuits and electro-optical modulators (EOM). The
sum (+) and difference (--) of the photocurrents measured by HD$_{1}$\ and HD%
$_{2}$\ are only used for the single-mode squeezing gate. When Fourier
transformation is implemented, they are not utilized. The resultant optical
mode is measured by the third homodyne detection system HD$_{3}$.

In the standard CV quantum teleportation process \cite{Furusawa1998}, the
amplitude and phase quadratures of output modes from BS$_{1}$ are measured
by two homodyne detection systems, the measurement angles of which are
chosen as 0 and $\pi /2$, respectively. However, in the presented quantum
logic operation the measurement angle will be chosen arbitrarily, and the
squeezing degree of the squeezing gate will be determined by the measurement
angle. Thus, we can say that the CVQC logic operation is implemented by
means of a\ CV quantum teleportation process with an arbitrarily chosen
measurement angle.

If the input mode is coupled to mode E$_{1}$\ with a $\pi /2$\ phase
difference on BS$_{1}$, the measurement results of HD$_{1}$\ and HD$_{2}$, $%
\hat{x}_{d1}$\ and $\hat{x}_{d2}$, are expressed by 
\begin{eqnarray}
\hat{x}_{d1} &=&\frac{\cos \theta _{1}(\hat{x}_{in}-\hat{p}_{1})+\sin \theta
_{1}(\hat{p}_{in}+\hat{x}_{1})}{\sqrt{2}}, \\
\hat{x}_{d2} &=&\frac{\cos \theta _{2}(\hat{x}_{in}+\hat{p}_{1})+\sin \theta
_{2}(\hat{p}_{in}-\hat{x}_{1})}{\sqrt{2}},  \notag
\end{eqnarray}%
where $\theta _{1}$\ and $\theta _{2}$\ are the measurement angles of HD$_{1}
$\ and HD$_{2}$, respectively. Choosing $\theta _{2}=-\theta _{1}$, the
amplitude and phase quadratures of the resultant mode equal to: 
\begin{eqnarray}
\dbinom{\hat{x}_{out}}{\hat{p}_{out}} &=&\dbinom{\hat{x}_{2}}{\hat{p}_{2}}%
+G_{S}\dbinom{\hat{x}_{d1}}{\hat{x}_{d2}}  \notag \\
&=&%
\begin{pmatrix}
\cot \theta _{1} & 0 \\ 
0 & \tan \theta _{1}%
\end{pmatrix}%
\dbinom{\hat{x}_{in}}{\hat{p}_{in}}+\dbinom{\hat{\delta}_{1}}{-\hat{\delta}%
_{2}},  \label{3}
\end{eqnarray}%
where 
\begin{equation}
G_{S}=%
\begin{pmatrix}
\frac{1}{\sqrt{2}\sin \theta _{1}} & \frac{1}{\sqrt{2}\sin \theta _{1}} \\ 
\frac{1}{\sqrt{2}\cos \theta _{1}} & \frac{-1}{\sqrt{2}\cos \theta _{1}}%
\end{pmatrix}%
\end{equation}%
\textbf{\ }is the corresponding gain factor and\textbf{\ }$\hat{\delta}_{1}=%
\hat{x}_{1}+\hat{x}_{2}$\ and $\hat{\delta}_{2}=\hat{p}_{1}-\hat{p}_{2}$\
are the excess noises of the amplitude and phase quadratures of the EPR
entangled state\ respectively, which result from the imperfect entanglement
of the resource state and whose variances depend on the squeezing parameter $%
r_{E}$\ of the EPR state by $\left\langle \Delta ^{2}(\hat{x}_{1}+\hat{x}%
_{2})\right\rangle =\left\langle \Delta ^{2}(\hat{p}_{1}-\hat{p}%
_{2})\right\rangle =e^{-2r_{E}}/2$. For an ideal EPR state $r_{E}\rightarrow
\infty $\ and thus $\hat{\delta}_{1}=\hat{\delta}_{2}=0$. The ideal EPR
state does not exist really since it requires infinite energy \cite%
{Menicucci2006}.

Comparing Eq. (3) in the case of ideal EPR state with Eq. (1), we can see
that the transformation corresponds\ to a single-mode amplitude and phase
squeezing gate with $\cot \theta _{1}=e^{-r}$\ and $e^{r}$, respectively. In
this case, the transformation matrix is given by 
\begin{equation}
\mathbf{S=}%
\begin{pmatrix}
\cot \theta _{1} & 0 \\ 
0 & \tan \theta _{1}%
\end{pmatrix}%
.
\end{equation}%
Equation (5) shows that the squeezing parameter\ $r$\ depends on\ the
measurement angles. When the measurement angle is varied from $45%
{{}^\circ}%
$\ to $0%
{{}^\circ}%
$\ the squeezing degree of the squeezing gate increases from\textbf{\ }0 to $%
-\infty $.\textbf{\ }The squeezing level can be controlled by choosing
different measurement angles. The measurement angles ($\theta _{1},\theta
_{2}$) for the squeezing levels of $-4$, $-8$,\ and $-12$\ dB are ($32.25%
{{}^\circ}%
,-32.25%
{{}^\circ}%
$), ($21.70%
{{}^\circ}%
,-21.70%
{{}^\circ}%
$), and ($14.10%
{{}^\circ}%
,-14.10%
{{}^\circ}%
$), respectively.

When we take $\theta _{1}$\ $=0$\ and $\theta _{2}=-\pi /2$, the amplitude
and phase quadratures of the resultant mode are 
\begin{eqnarray}
\dbinom{\hat{x}_{out}}{\hat{p}_{out}} &=&\dbinom{\hat{x}_{2}}{\hat{p}_{2}}%
+G_{F}\dbinom{\hat{x}_{d1}}{\hat{x}_{d2}}  \notag \\
&=&\mathbf{F}\dbinom{\hat{x}_{in}}{\hat{p}_{in}}+\dbinom{\hat{\delta}_{1}}{-%
\hat{\delta}_{2}},
\end{eqnarray}%
where 
\begin{equation}
G_{F}=%
\begin{pmatrix}
0 & \sqrt{2} \\ 
\sqrt{2} & 0%
\end{pmatrix}%
\end{equation}%
is the corresponding gain factor of the feedforward circuit. The
transformation matrix $\mathbf{F}=%
\begin{pmatrix}
0 & -1 \\ 
1 & 0%
\end{pmatrix}%
$\ just corresponds to a Fourier transformation. Thus, a Fourier
transformation operation can also be implemented with the experimental
system of Fig. 1 (b) only by choosing appropriate measurement angles and
feedforward circuit.

In the one-way quantum computation scheme with the four-mode cluster state
as the resource , the excess noises of the amplitude ($\hat{\delta}_{xc}$)
and phase ($\hat{\delta}_{pc}$) quadratures of the output mode for the
squeezing of $a$\ dB ($a<0$\ and $a>0$\ correspond to phase squeezing and
amplitude squeezing, respectively) are given by \cite{Ukai2011} 
\begin{equation}
\dbinom{\hat{\delta}_{xc}}{\hat{\delta}_{pc}}=\dbinom{\frac{1}{\sqrt{2}}%
e^{-r_{c}}\hat{p}_{1}^{0}-\sqrt{\frac{5}{2}}e^{-r_{c}}\hat{p}_{2}^{0}}{-%
\sqrt{\frac{5}{2}}e^{-r_{c}}\hat{p}_{3}^{0}+\frac{1}{\sqrt{2}}e^{-r_{c}}\hat{%
p}_{4}^{0}}\qquad   \label{s}
\end{equation}%
when $V=10^{a/10}\leq \frac{3}{2}$, and 
\begin{equation}
\dbinom{\hat{\delta}_{xc}}{\hat{\delta}_{pc}}=\dbinom{\frac{e^{-r_{c}}\left[
3\hat{p}_{1}^{0}/V-2\sqrt{5}\hat{p}_{2}^{0}+\sqrt{2V-3}\left( \sqrt{5}\hat{p}%
_{3}^{0}+\hat{p}_{4}^{0}\right) /V\right] }{2\sqrt{2}}}{\frac{e^{-r_{c}}%
\left[ \sqrt{2V-3}\hat{p}_{1}^{0}-\sqrt{5}\hat{p}_{3}^{0}+\hat{p}_{4}^{0}%
\right] }{\sqrt{2}}}  \label{xs}
\end{equation}%
when $V>\frac{3}{2}$, where $r_{c}$\ is the squeezing parameter of four
phase squeezed state $\hat{p}_{1-4}$,\ and the superscript 0 represents the
vacuum mode. From Eq. (\ref{s}) we can calculate the variance of the excess
noise for phase squeezing: $\left\langle \Delta ^{2}\hat{\delta}%
_{xc}\right\rangle =\left\langle \Delta ^{2}\hat{\delta}_{pc}\right\rangle
=3e^{-2r_{c}}/4$, where the noise variance of the vacuum mode is normalized
to $\left\langle \Delta ^{2}\hat{x}^{0}\right\rangle =\left\langle \Delta
^{2}\hat{p}^{0}\right\rangle =1/4$. In our scheme the variance of the excess
noise is the quantum correlation variances of the EPR entangled state \cite%
{Wang20102}, i.e. $\left\langle \Delta ^{2}(\hat{x}_{1}+\hat{x}%
_{2})\right\rangle =\left\langle \Delta ^{2}(\hat{p}_{1}-\hat{p}%
_{2})\right\rangle =e^{-2r_{E}}/2$. Comparing the two cases, we find that
the variances of the excess noises in the scheme using the EPR resource
state is 2/3 of that using a four-mode cluster state if the squeezing degree
of the initial resource squeezed state is the same.

\begin{figure}[tbp]
\begin{center}
\setlength{\belowcaptionskip}{-3pt} 
\includegraphics[width=80mm]{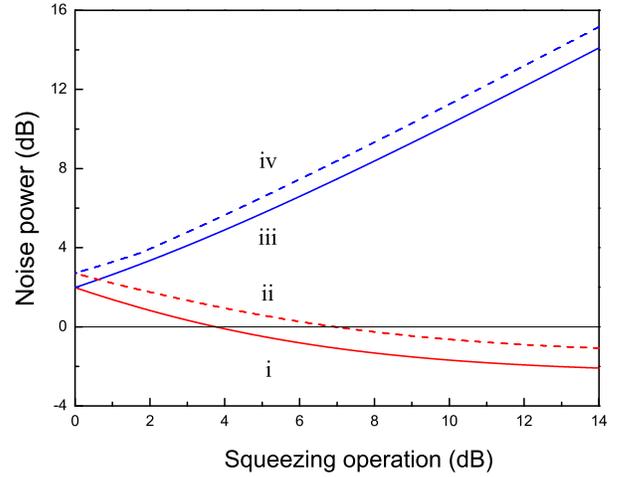} 
\end{center}
\caption{(Color online) The dependence of the noise power of the output mode
on the amplitude squeezing level of the squeezing operation for different
resource states. Input state: a vacuum state. Traces i and iii (solid lines)
correspond to squeezed and antisqueezed noises using an EPR entangled state
as a resource state, respectively. Traces ii and iv (dashed lines)
correspond to squeezed and antisqueezed noises with a four-mode cluster
state as resource state, respectively. The initial resource squeezing is $%
-5.3$ dB for the two cases.}
\end{figure}

Fig. 2 compares the noise powers of the output modes of the amplitude
squeezing operation implemented in the two systems using the EPR entangled
state (solid lines) and the four-mode cluster state (dashed lines) as
resource states, in which the initial squeezing of the resource states is
taken to be the same ($-5.3$ dB). The noise power is calculated by $10\log
_{10}[B/B_{0}]$ dB, where $B$ represents the noise variance of the
quadrature component and $B_{0}=1/4$ is the normalized vacuum noise. In this
case, 0 dB in Fig. 2 corresponds to the vacuum noise level. It is obvious
that both squeezed (traces i and ii) and antisqueezed (traces iii and iv)
noise powers of the output modes obtained by the system using the EPR
entangled state are lower than those obtained by the system using the
four-mode cluster state in Ref. \cite{Ukai2011}. Therefore, for a given
initial squeezing resource, the squeezing gate based on EPR entanglement can
generate the squeezed states with a higher squeezing degree and lower
antisqueezing noises than that obtained using the four-mode cluster state.

\section{Experimental set-up and results}

\subsection{\textbf{Experimental set-up}}

The experimental set-up is shown in Fig. 1 (b). The non-degenerate optical
parametric amplifier (NOPA) is pumped by a continuous wave intra-cavity
frequency-doubled and frequency-stabilized Nd:YAP-LBO (Nd-doped YAlO$_{3}$\
perovskite-lithium triborate) laser with two output wavelengths at 540 nm
and 1080 nm \cite{WangIEEE2010}. The NOPA consists of an $\alpha $-cut
type-II potassium titanyl phosphate (KTP) crystal and a concave mirror \cite%
{Wang20102}. The front face of the KTP crystal is coated to be used for the
input coupler and the concave mirror serves as the output coupler of the
squeezed states. The transmissions of the input (output) coupler at 540 and
1080 nm are $99.8\%$ ($0.5\%$) and $0.04\%$ ($5.2\%$), respectively. The EPR
entangled states at 1080 nm are generated via the frequency-down-conversion
process of the pump field at 540 nm inside the NOPA. The amplitude
anti-correlated ($\hat{x}_{1}+\hat{x}_{2}\rightarrow 0$) and phase
correlated ($\hat{p}_{1}-\hat{p}_{2}\rightarrow 0$) EPR entangled optical
beams are obtained when the NOPA is operated at the deamplification
condition, which corresponds to locking the relative phase between the pump
laser and the injected signal to $(2n+1)\pi $ ($n$ is the integer) \cite%
{Wang20102}. The experimentally measured squeezing of the EPR entangled
state is about $-4.0$ dB.

\subsection{Squeezing operation}

\begin{figure*}[tbp]
\begin{center}
\setlength{\belowcaptionskip}{-3pt} 
\includegraphics[width=160mm]{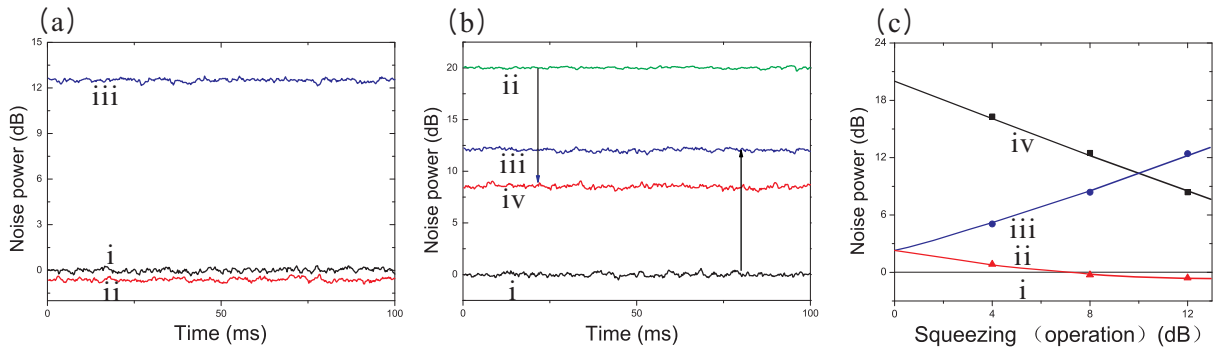} 
\end{center}
\caption{(Color online) The experimental results of the single-mode
squeezing operation. (a): the $-12$ dB squeezing operation with a vacuum
input state, (b): $-12$ dB squeezing operation with a $\hat{p}$-coherent
state. Trace i: SNL, traces ii: input variances of the $\hat{p}$-coherent
state, traces iii and iv: anti-squeezing and squeezing noise. (c):
experimental results (dots) and theoretical curves (lines) for $-4$, $-8$
and $-12$ dB squeezing operation. Traces ii and iii: squeezing and
anti-squeezing with a vacuum state, trace iv: squeezing for a $\hat{p}$%
-coherent state. Measurement frequency: 2 MHz, parameters of the spectrum
analyzer: resolution bandwidth: 30kHz, video bandwidth: 100Hz.}
\end{figure*}

Fig. 3 (a) and (b) show the output noise power of the $-12$ dB phase
squeezing operation with a vacuum input and a $\hat{p}$-coherent input,
respectively. Trace i (black line) is the shot-noise-level (SNL); traces ii
and iii (red and blue lines) are the squeezed and anti-squeezed noises,
respectively. Although in the ideal case with $\hat{\delta}_{1}=\hat{\delta}%
_{2}=0$, the input vacuum state should be squeezed $-12$ dB, in the
practical experiment the input vacuum mode is squeezed 0.6 dB below the
corresponding SNL due to\textbf{\ }the influence of the excess noises
introduced by the imperfect EPR entanglement.

In order to test the generality of the squeezing operation, we implement a
squeezing operation on a $\hat{p}$-coherent input state with a modulation
signal of 20 dB on its phase quadratures. In Fig. 3 (b), trace ii (green
line) stands for the input coherent state. The squeezing (trace iv) and
anti-squeezing (trace iii) noise levels of the output mode are 8.2 dB and
12.2 dB above the SNL, respectively. Fig. 3 (c) shows the three different
squeezing levels ($-4$, $-8$, and $-12$ dB) with a vacuum state (trace ii
and iii) and a $\hat{p}$-coherent state (trace iv)\ as input states,
respectively. The measurement results agree well with the theoretical curves
(solid lines).

Besides the squeezed noise level of the output mode, we also use the
fidelity $F=\left\{ \text{Tr}[(\sqrt{\hat{\rho}_{1}}\hat{\rho}_{2}\sqrt{\hat{%
\rho}_{1}})^{1/2}]\right\} ^{2}$, which denotes the overlap between the
experimentally obtained output state $\hat{\rho}_{2}$\ and the ideal output
sate $\hat{\rho}_{1}$, to quantify the performance of the squeezing
operation. The fidelity for two Gaussian states $\hat{\rho}_{1}$\ and $\hat{%
\rho}_{2}$\ with covariance matrices $\mathbf{A}_{j}$\ and mean amplitudes $%
\mathbf{\alpha }_{j}\equiv (\alpha _{jx},\alpha _{jp})$\ ($j=1,2$)\ is
expressed as \cite{Nha2005,Scutaru1998} 
\begin{equation}
F=\frac{2}{\sqrt{\Delta +\sigma }-\sqrt{\sigma }}\exp [-\mathbf{\beta }^{T}(%
\mathbf{A}_{1}+\mathbf{A}_{2})^{-1}\mathbf{\beta }],
\end{equation}%
\ where $\Delta =\det (\mathbf{A}_{1}+\mathbf{A}_{2}),$\ $\sigma =(\det 
\mathbf{A}_{1}-1)(\det \mathbf{A}_{2}-1),$\ $\mathbf{\beta }=\mathbf{\alpha }%
_{2}-\mathbf{\alpha }_{1},$ and $\mathbf{A}_{1}$ and $\mathbf{A}_{2}$ are
for the ideal ($\hat{\rho}_{1}$) and experimental ($\hat{\rho}_{2}$) output
states, respectively. The covariance matrices $\mathbf{A}_{j}$\ ($j=1,2$)
for the target mode are given by 
\begin{eqnarray}
\mathbf{A}_{out1} &=&4\left[ 
\begin{array}{cc}
\left\langle \Delta ^{2}\hat{x}_{out}\right\rangle _{1} & 0 \\ 
0 & \left\langle \Delta ^{2}\hat{p}_{out}\right\rangle _{1}%
\end{array}%
\right] ,\qquad  \\
\mathbf{A}_{out2} &=&4\left[ 
\begin{array}{cc}
\left\langle \Delta ^{2}\hat{x}_{out}\right\rangle _{2} & 0 \\ 
0 & \left\langle \Delta ^{2}\hat{p}_{out}\right\rangle _{2}%
\end{array}%
\right] .
\end{eqnarray}%
The coefficient 4\ comes from the normalization of the SNL. Since the noise
of a vacuum state is defined as 1/4 above, while in the fidelity formula the
vacuum noise is normalized to 1, a coefficient 4\ appears in the expressions
of covariance matrices.\textbf{\ }In the case of infinite squeezing, the
fidelity for the output state equals 1, which can be calculated from Eq. (%
\ref{3})\ with $\hat{\delta}_{1}=\hat{\delta}_{2}=0$ ($r\rightarrow \infty $%
).

\begin{figure}[tbp]
\begin{center}
\setlength{\belowcaptionskip}{-3pt} 
\includegraphics[width=80mm]{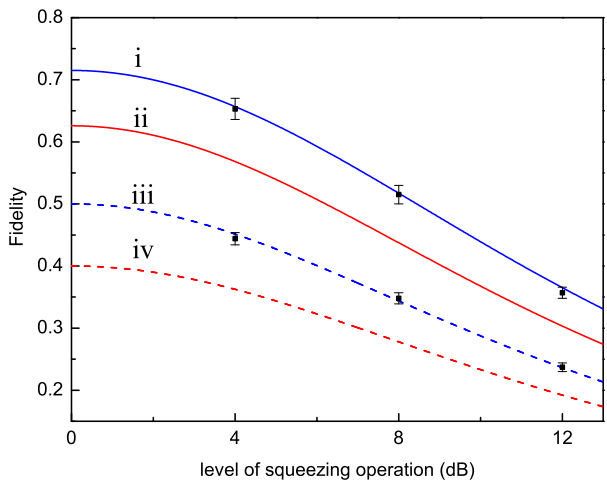} 
\end{center}
\caption{(Color online) The fidelity as a function of phase-squeezing.
Traces i and iii are fidelities with and without EPR entanglement as a
resource, respectively. Traces ii and iv are fidelities with and without a
four-mode cluster state as a resource, respectively. The initial resource
squeezing is $-4.0$ dB for traces i and ii.}
\end{figure}

Fig. 4 shows the fidelity as a function of the phase squeezing. We can see
that the fidelity with the $-4.0$ dB EPR state as a resource state (trace i,
blue solid line) is higher than the classical limit which is obtained by
using the coherent state to substitute for the EPR state (trace iii, blue
dashed line). For the comparison, we calculate the fidelity based on the
four-mode cluster state with the same initial squeezing resource of $-4.0$
dB.\textbf{\ }Traces ii (red line) and iv (red dash) are the fidelities with
and without four-mode cluster state ($-4.0$\ dB initial squeezing) as a
resource state. The fidelity of squeezing operation using the EPR state as a
resource state is higher than that using the four-mode cluster state. This
is because the excess noise deriving from the squeezing operation in the
scheme using the EPR state is only 2/3 of that based on the four-mode
cluster state \cite{Ukai2011}. Experimentally measured data are marked on
the graph with black dots, which are in good agreement with the theoretical
expectation.

\subsection{Fourier operation}

\begin{figure}[tbp]
\begin{center}
\setlength{\belowcaptionskip}{-3pt} \includegraphics[width=85mm]{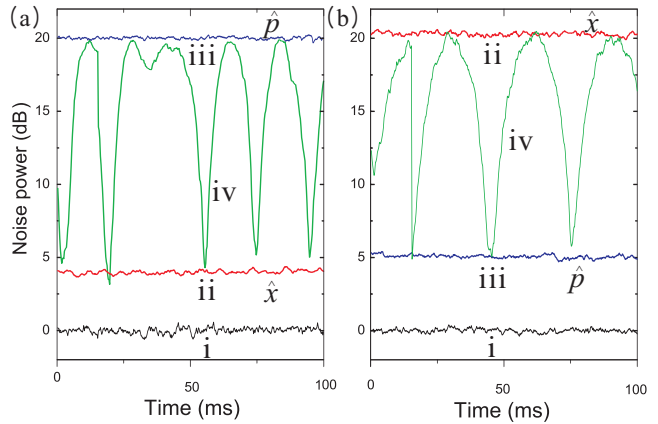}
\end{center}
\caption{(Color online) The experimental results of Fourier transformation.
(a): input state, (b): output state. Trace i: SNL, traces ii and iii:
amplitude and phase quadratures, respectively. Trace iv: noise power when
the phase of the homodyne detection system is scanned. Measurement
frequency: 2 MHz, parameters of the spectrum analyzer: resolution bandwidth:
30kHz, video bandwidth: 100Hz.}
\end{figure}

It has been theoretically proved in section II that when the measurement
angles of HD$_{1}$\ and HD$_{2}$\ are taken as 0 and $\pi /2$, respectively,
the input mode will complete the Fourier transformation via a teleportation
process in the experimental system of Fig. 1(b). Fig. 5 shows the
experimental results of Fourier transformation with a coherent input.
Figures 5(a) and 5(b) correspond to noise powers of the input and output
states, respectively. Trace i (black line) is the SNL, and traces ii and iii
(red and blue lines) stand for the average noise levels of the amplitude and
phase quadratures of the input [Fig.5(a)] and output [Fig. 5(b)] modes,
respectively. Trace iv (green line) is the noise power spectrum of the input
[Fig. 5(a)] and output [Fig. 5(b)] states measured by scanning the phase of
the homodyne detection system. A coherent state with a 4 dB amplitude
modulation signal\ on the amplitude quadrature and a 20 dB amplitude
modulation signal on the phase quadrature is used for the input state [Fig.5
(a)].\textbf{\ }Fig. 5 (b) shows the amplitude and phase quadratures of the
output state after the Fourier transformation. Comparing Figures 5(a) and
5(b), we can see that the input mode has been rotated 90$%
{{}^\circ}%
$\ in the phase space, and thus, the Fourier transformation from the phase
(amplitude) quadrature to the amplitude (phase) quadrature has been achieved.

\section{Conclusion}

In conclusion, we have designed and experimentally demonstrated two
essential one-mode LUBO transformations based on the use of an EPR entangled
state. Squeezing and Fourier transformation operations are implemented on an
experimental set-up. These operations are easily controlled by adjusting the
phase of the local oscillator in the homodyne detectors. The calculation
accuracy of one-way CVQC depends on the initial resource squeezing since an
imperfect resource state will introduce excess noises into the calculated
resultant states via the gate operations. The excess noises deriving from
the EPR system are less than those from the four-mode cluster system, so
better accuracy can be obtained by the gates using EPR entanglement under
the condition of applying the same initial squeezing resource.

Finally, we demonstrate theoretically that the presented experimental set-up
can also complete a cascaded single-mode logic operation consisting of a
squeezing operation and a Fourier transformation, which shows further the
versatility of the system. If the phase difference\textbf{\ }between\ the
input mode and a submode of the EPR entangled state on BS$_{1}$ is taken as
zero, the measurement results from the two homodyne detection systems will
be 
\begin{eqnarray}
\hat{x}_{d1} &=&\frac{\cos \theta _{1}(\hat{x}_{in}-\hat{x}_{1})+\sin \theta
_{1}(\hat{p}_{in}-\hat{p}_{1})}{\sqrt{2}}, \\
\hat{x}_{d2} &=&\frac{\cos \theta _{2}(\hat{x}_{in}+\hat{x}_{1})+\sin \theta
_{2}(\hat{p}_{in}+\hat{p}_{1})}{\sqrt{2}}.  \notag
\end{eqnarray}%
Choosing $\theta _{2}=-\theta _{1},$\ the quadrature components of the
output mode equal to 
\begin{eqnarray}
\dbinom{\hat{x}_{out}}{\hat{p}_{out}} &=&\dbinom{\hat{x}_{2}}{\hat{p}_{2}}%
+G_{FS}\dbinom{\hat{x}_{d1}}{\hat{x}_{d2}}  \label{fs} \\
&=&%
\begin{pmatrix}
0 & -1 \\ 
1 & 0%
\end{pmatrix}%
\begin{pmatrix}
\cot \theta _{1} & 0 \\ 
0 & \tan \theta _{1}%
\end{pmatrix}%
\dbinom{\hat{x}_{in}}{\hat{p}_{in}}+\dbinom{\hat{\delta}_{1}}{-\hat{\delta}%
_{2}},  \notag
\end{eqnarray}%
which stands for a cascaded squeezing operation followed by a Fourier
transformation,\textbf{\ }where 
\begin{equation}
G_{FS}=%
\begin{pmatrix}
\frac{-1}{\sqrt{2}\cos \theta _{1}} & \frac{1}{\sqrt{2}\cos \theta _{1}} \\ 
\frac{1}{\sqrt{2}\sin \theta _{1}} & \frac{1}{\sqrt{2}\sin \theta _{1}}%
\end{pmatrix}%
\end{equation}%
is the corresponding gain factor. Eq. (\ref{fs}) can also be written as 
\begin{equation}
\dbinom{\hat{x}_{out}}{\hat{p}_{out}}=%
\begin{pmatrix}
-\tan \theta _{1} & 0 \\ 
0 & \cot \theta _{1}%
\end{pmatrix}%
\dbinom{\hat{x}_{in}}{\hat{p}_{in}}+\dbinom{\hat{\delta}_{1}}{-\hat{\delta}%
_{2}}.
\end{equation}%
Equation (16) means that the squeezing operation followed by a Fourier
transformation is equivalent to rotating the measurement angle of the
homodyne detection in the squeezing gate by $90%
{{}^\circ}%
$,\ and thus, the two operations can be achieved in one step.\textbf{\ }The
excess noise induced by imperfect resource squeezing only equals to that of
a squeezing gate.

Although two essential single-mode LUBO transformations have been realized
by the EPR system, to implement high-order and universal one-way QC,
large-scale cluster states and additional non-Gaussian operations are
required \cite{Menicucci2006}. However, the presented schemes can be
utilized as the basic modules in a full quantum computer using CV cluster
entanglement. Saving quantum resources and decreasing excess noise are two
favorite features of the EPR system for building a practicable one-way
quantum computer with continuous quantum variables of optical modes.

\section*{Acknowledgments}

This research was supported by the National Basic Research Program of China
(Grant No. 2010CB923103), NSFC (Grant Nos. 11174188, 61121064), OIT and
Shanxi Scholarship Council of China (Grant No. 2012-010).


\begin{thebibliography}{99}
\bibitem{Nielsen2000} M. A. Nielsen and I. L. Chuang, \textit{Quantum
computation and Quantum information} (Cambridge University Press, Cambridge,
2000).

\bibitem{Loock1} A. Furusawa and P. van Loock, \textit{Quantum Teleportation
and Entanglement} (Wiley-VCH, Berlin, 2011).

\bibitem{Raussendorf2001} R. Raussendorf, and H. J. A. Briegel, Phys. Rev.
Lett. \textbf{86}, 5188-5191 (2001).

\bibitem{Walther2005} P. Walther, K. J. Resch, T. Rudolph, E. Schenck, H.
Weinfurter, V. Vedral, M. Aspelmeyer, and A. Zeilinger, Nature, \textbf{434}%
, 169 (2005).

\bibitem{Prevedel2007} R. Prevedel, P. Walther, F. Tiefenbacher, P. B\"{o}%
hi, R. Kaltenbaek, T. Jennewein and A. Zeilinger, Nature, \textbf{445}, 65
(2007).

\bibitem{Chen2007} K. Chen, C. M. Li, Q. Zhang, Y. A. Chen, A. Goebel, S.
Chen, A. Mair, and J. W. Pan, Phys. Rev. Lett. \textbf{99}, 120503 (2007).

\bibitem{Menicucci2006} N. C. Menicucci, P. van Loock, M. Gu, C. Weedbrook,
T. C. Ralph, and M. A. Nielsen, Phys. Rev. Lett. \textbf{97}, 110501 (2006).

\bibitem{Zhang2006} J. Zhang, and S. L. Braunstein, Phys. Rev. A \textbf{73}%
, 032318 (2006).

\bibitem{Loock2007} P. van Loock, C. Weedbrook and M. Gu, Phys. Rev. A 
\textbf{76}, 032321 (2007).

\bibitem{Su2007} X. Su, A. Tan, X. Jia, J. Zhang, C. Xie, and K. Peng, Phys.
Rev. Lett. \textbf{98}, 070502 (2007).

\bibitem{Yukawa2008} M. Yukawa, R. Ukai, P. van Loock, and A. Furusawa,
Phys. Rev. A \textbf{78}, 012301 (2008).

\bibitem{Weedbrook2012} C. Weedbrook, S. Pirandola, R. Garc\'{\i}a-Patr\'{o}%
n, N. J. Cerf, T. C. Ralph, J. H. Shapiro, and S. Lloyd, Rev. Mod. Phys. 
\textbf{84}, 621 (2012).

\bibitem{Ukai2010} R. Ukai, J. I. Yoshikawa, N. Iwata, P. van Loock, and A.
Furusawa, Phys. Rev. A \textbf{81}, 032315 (2010).

\bibitem{Ukai2011} R. Ukai, N. Iwata, Y. Shimokawa, S. C. Armstrong, A.
Politi, J. I. Yoshikawa, P. van Loock, and A. Furusawa, Phys. Rev. Lett. 
\textbf{106}, 240504 (2011).

\bibitem{Zwi} M. Zwierz, C. A. P\'{e}rez-Delgado, and P. Kok, Phys. Rev. A 
\textbf{82}, 042320 (2010).

\bibitem{Yoshikawa2008} J. I. Yoshikawa, Y. Miwa, A. Huck, U.~L. Andersen,
P. van Loock, and A. Furusawa, Phys. Rev. Lett. \textbf{101}, 250501 (2008).

\bibitem{Miwa2009} Y. Miwa, J. I. Yoshikawa, P. van Loock, and A. Furusawa,
Phys. Rev. A \textbf{80}, 050303(R) (2009).

\bibitem{Wang2010} Y. Wang, X. Su, H. Shen, A. Tan, C. Xie, and K. Peng,
Phys. Rev. A \textbf{81}, 022311 (2010).

\bibitem{Ukai20112} R. Ukai, S. Yokoyama, J. I. Yoshikawa, P. van Loock, and
A. Furusawa, Phys. Rev. Lett. \textbf{107}, 250501 (2011).

\bibitem{Furusawa1998} A. Furusawa, J. L. Sorenson, S. L. Braunstein, C. A.
Fuchs, H. J. Kimble, and E. S. Polzik, Science \textbf{282}, 706 (1998).

\bibitem{Wang20102} Y. Wang, H. Shen, X. Jin, X. Su, C. Xie, and K. Peng,
Opt. Express \textbf{18}, 6149-6155 (2010).

\bibitem{WangIEEE2010} Y. Wang, Y. Zheng, C. Xie, and K. Peng, IEEE J.
Quantum Electronics \textbf{47}, (7), 1006 (2011).

\bibitem{Nha2005} H. Nha, and H. J. Carmichael, Phys. Rev. A \textbf{71},
032336 (2005).

\bibitem{Scutaru1998} H. Scutaru, J. Phys. A \textbf{31}, 3659 (1998).
\end{thebibliography}
\end{document}